\documentclass[universe,accept,article,pdfpaper,10pt,a4paper]{Definitions/mdpi} 
\usepackage{hyperref}%[breaklinks]
\usepackage[english]{babel}
\usepackage[utf8]{inputenc}
\usepackage[OT2,T1]{fontenc}
\usepackage{graphicx}
\usepackage{color}
\usepackage{amsmath,amssymb,amsfonts,amsthm}
\usepackage{esint}
\usepackage{bm,bbm}

\newcommand{\be}{\begin{equation}}
\newcommand{\ee}{\end{equation}}
\newcommand{\ba}{\begin{eqnarray}}
\newcommand{\ea}{\end{eqnarray}}
\def\bs{\begin{subequations}}
\def\es{\end{subequations}}

\renewcommand{\geq}{\geqslant}
\def\a{\alpha}
\def\b{\beta}
\def\de{\delta}
\def\g{\gamma}
\def\la{\lambda}

\def\om{\omega}

\def\N{\nabla}

\def\cF{\mathcal{F}}
\def\cG{\mathcal{G}}

\def\cK{\mathcal{K}}
\def\cL{\mathcal{L}}

\def\cS{\mathcal{S}}

\def\p{\partial}

\def\B{\Box}
\newcommand{\Eq}[1]{(\ref{#1})}

\def\cob{\color{blue}}

\newcommand{\oarX}[1]{\href{http://arxiv.org/abs/#1}{{\ttfamily\cob arXiv:#1}}}
\newcommand{\arX}[1]{\href{http://arxiv.org/abs/#1}{{\ttfamily\cob arXiv:#1}}}
\newcommand{\doin}[6]{\href{http://dx.doi.org/#1}{{\cob {\it #2 #3} {\bf #6}, {\it #4}, #5}}}
\newcommand{\doinn}[5]{\href{http://dx.doi.org/#1}{{\cob {\it #2} {\bf #5}, {\it #3}, #4}}}
\newcommand{\doij}[5]{\href{http://dx.doi.org/#1}{{\cob {\it #2} {\bf #5}, {\it #3}, #4}}}

\newcommand{\ndoinn}[5]{\href{#1}{{\cob {\it #2} {\bf #5}, {\it #3}, #4}}}
\newcommand{\tia}[1]{#1.}

\def\rme{e}
\def\rmd{d}
\def\rmi{i}

\newcounter{listcounter}

\newcommand\cyr{%
\renewcommand\rmdefault{wncyr}%
\renewcommand\sfdefault{wncyss}%
\renewcommand\encodingdefault{OT2}%
\normalfont
\selectfont}
\DeclareTextFontCommand{\textcyr}{\cyr}

\firstpage{1} 
\makeatletter 
\articlenumber{95}
\doinum{10.3390/universe4090095}
\pubvolume{4}
\pubyear{2018}
\copyrightyear{2018}
\history{Date: June 25, 2018}

\Title{Taming the beast: diffusion method in nonlocal gravity}

\Author{Gianluca Calcagni}

\AuthorNames{Gianluca Calcagni}

\address{Instituto de Estructura de la Materia, CSIC, Serrano 121, 28006 Madrid, Spain; g.calcagni@csic.es}
\firstnote{This paper is based on a talk at the International Conference on Quantum Gravity, Shenzhen, China, 26–28 March 2018.}

\corres{Correspondence: g.calcagni@csic.es}

\abstract{We present a method to solve the nonlinear dynamical equations of motion in gravitational theories with fundamental nonlocalities of a certain type. For these specific form factors, which appear in some renormalizable theories, the number of field degrees of freedom and of initial conditions is finite.}


\begin{document}

%%%%%%%%%%%%%%%%%%%%%%%%%%%%%%%%%%%%%%%%%%
%%%%%%%%%%%%%%%%%%%%%%%%%%%%%%%%%%%%%%%%%%

\section{Introduction}

``Nonlocal quantum gravity'' is an umbrella name including at least two different settings. The first group, which includes general relativity, consists in classically local gravitational theories which receive quantum corrections such that the effective one-loop action is nonlocal (i.e., it is made of operators with infinitely many derivatives). Examples are the approaches by Woodard and Maggiore (and collaborators) \cite{DeWo1,DeWo2,TW,MaMa,BDFM}. In this brief review of recent results, we will confine ourselves to the second meaning of the term, where nonlocality is fundamentally present already at the classical level and, thanks to the suppression of the graviton propagator in the ultraviolet (UV), the theory is renormalizable or finite. Its history runs on two parallel tracks. In one, mainly developed with discontinuous effort from the 1970s to the 1990s by Alebastrov, Efimov, Krasnikov, Kuz'min, Moffat and Tomboulis among others \cite{AE1,AE2,Efi77,Kra87,Kuz89,Tom97}, nonlocal operators were early recognized as an opportunity to improve the renormalization properties of scalar \cite{AE1,AE2,Efi77} and gauge \cite{Kra87,Mof1,Tom97} field theories, with some interest also in gravity \cite{Kuz89,Tom97}. In another (from the late 1980s to the early 2000s), the same operators made their appearance in string theory as nonperturbative effects in the low-energy limit \cite{W2,KS1,EW1,KS2,AMZ2,BF2,Tse95,Sie03,AKBM,MoZ,AJK,Ohm03}. Nowadays, nonlocal quantum gravity has achieved a high degree of independence both from these antecedents and from other proposals, to the point where it can be considered as one of the most promising and accessible candidates for a theory where the gravitational force consistently obeys the laws of quantum mechanics. In particular, there exist several renormalization results, both at finite order and to all orders in a perturbative Feynman-diagram expansion, which showed that the good UV properties guessed at the level of power-counting indeed hold rigorously (e.g., \cite{MoRa1,MoRa2}).

Despite the investment of much effort in taming fundamental nonlocality, several questions remain open to date:
\begin{enumerate}
\item Is the Cauchy problem well-defined?
\item If so, how many initial conditions must one specify for a solution?
\item How many degrees of freedom are there?
\item How to construct nontrivial solutions?
\item Is causality violated?
\item Are singularities resolved at the classical or quantum level?
\end{enumerate}
Here we will give the following answers to some of these issues:
\begin{enumerate}
\item Yes, for the form factors appearing in fundamental theories (not for all conceivable form factors).
\item Two or higher for a scalar field theory and four or higher for gravity (depending on the form factor), but finite.
\item Eight (in $D=4$ dimensions).
\item Via the diffusion method.
\end{enumerate}
Let us now examine where these cryptic responses come from. The main results can be found in \cite{CMN2,CMN3}, which we review and expand on some points. In Section \ref{disc}, we will comment on the physical applications of the method described here.

%%%%%%%%%%%%%%%%%%%%%%%%%%%%%%%%%%%%%%%%%%

\section{Action and form factors}

The classical fundamental (not effective) action of the theory is
\be
S = \frac{1}{2\kappa^2}\int \rmd^D x \sqrt{-g} \left[R+G_{\mu\nu}\gamma(\Box) R^{\mu\nu} \right]\,,\label{action}
\ee
where $G_{\mu\nu}$ is the Einstein tensor and $\gamma(\Box)$ is a weakly or quasi-polynomial nonlocal form factor. Formally, $\g$ is an analytic function that can be expressed as an infinite series with infinite convergence radius, $\gamma(\Box) = \sum_n c_n \Box^n$, although this expansion does not actually span the whole space of solutions in general \cite{cuta2}. As long as we require good properties at the quantum level (in particular, locality of the counterterms), the coefficients $c_n$ can be selected in a subclass of entire functions having special asymptotic properties \cite{Mod1,MoRa1,MoRa2,To15b,Kuz89,ALS,Tom97}. There are several form factors that preserve perturbative unitarity in \Eq{action} (see Table \ref{tab1}). In general, we can parametrized $\gamma$ as
\be\label{gamma}
\gamma(\Box) = \frac{\rme^{H(\Box)}-1}{\Box}\,,
\ee
where $H(\B)$ depends on the dimensionless combination $l^2\B$ and $l$ is a fixed length scale. The profile $H(\B)$ can be defined through the integral
\be
H(\B) := \a\int_0^{P(\B)} \rmd\omega\, \frac{1-f(\omega)}{\omega}\,,\label{Hz}
\ee
where $\a>0$ is real and $P(\B)$ is a generic function of $l^2\B$. The parameter $\a$ will not play any important role in what follows, but we included it to reproduce some form factors in the literature. Unless $P$ and $f$ are specified, $H(\B)$ is completely arbitrary. However, in a class of quantum gravities \cite{Mod1} $P(z)$ and $f(\om)$ must satisfy the following requirements: (i) $P(z)$ is a real polynomial $P_n(z)$ of degree $n$ with $P(0) = 0$; (ii) $f(\om)$ is an entire and real function on the real axis with $f(0) = 1$. A class of form factors of this type is generated by the function $f(\om)=\exp(-\om)$, so that
\be
\hspace{-0.5cm}
H(\B)=H^{\rm pol}(\B):=\a\left\{\ln P(\B)+\Gamma[0,P(\B)]+\gamma_{\rm E}\right\}\,,\label{FFgen}
\ee
where ${\rm Re}\,P(z)>0$, $\Gamma$ is the gamma function, and $\gamma_{\rm E}$ is the Euler--Mascheroni constant.\footnote{Generalizing to $f(\om)=\exp(-\om^n)$ yields the same expression with $\a\to\a/n$ and $P\to P^n$, but since we will keep $P$ generic we do not need this complication.} In particular, Kuz'min form factor \cite{Kuz89} corresponds to $P(\B)=P_1(\B):=-l^2\B$, while Tomboulis--Modesto form factor \cite{Tom97,Mod1} corresponds to a generic $P_{n}(\B)$ of order $n$ (derivative order $2n$). 

Another class of form factors is asymptotically exponential. In particular, the function $f(\om)=1-\om$ generates an exact monomial profile\footnote{Again, the generalization to $f(\om)=1-n\om^n$ is straightforward and not very instructive.} 
\be
H(\B)=H^{\rm exp}(\B):=\a P(\B)\,.\label{FFgen2}
\ee
If $\a=1$ and $P(\B)=P_1(\B)=-l^2\B$, we get the string-related exponential form factor \cite{BMS,CaMo2}, while if $\a=1$ and $P(\B)=P_2(\B)=l^4\B^2$ one obtains Krasnikov form factor \cite{Kra87} ($\a$ can be reabsorbed in the length and its value is not important). Quantum gravity with Kuz'min or Tomboulis--Modesto form factors is renormalizable \cite{Kuz89,Tom97,Mod1}; with the profile $H^{\rm exp}(\B)=P_1(\B)$, it is renormalizable if perturbative expansions with the resummed propagator are allowed \cite{TBM}; with Krasnikov profile $H^{\rm exp}(\B):=P_2(\B)$, it is believed to be renormalizable, although there is no complete proof \cite{Kra87}.
\begin{table}
\caption{Form factors in nonlocal gravity.\label{tab1}}
\begin{center}
\begin{tabular}{c|c|c|l}\hline
$H(\B)$ & $P(\B)$ & $f(\om)$ & Form factor name\\\hline\hline
\begin{tabular}{@{}c@{}}$H^{\rm pol}(\B):=\a\{\ln P(\B)+\Gamma[0,P(\B)]+\gamma_{\rm E}\}$ \end{tabular} & \begin{tabular}{@{}c@{}}$-l^2\B$ \\ $O(\B^n)$ \end{tabular} & $\rme^{-\om}$ & \begin{tabular}{@{}l@{}}Kuz'min \cite{Kuz89} \\ Tomboulis--Modesto \cite{Tom97,Mod1}\end{tabular}\\\hline
$H^{\rm exp}(\B):=\a P(\B)$ & \begin{tabular}{@{}c@{}}$-l^2\B$ \\ $l^4\B^2$\end{tabular} & $1-\om$
& \begin{tabular}{@{}l@{}}string-related \cite{BMS,CaMo2} \\ Krasnikov \cite{Kra87}\end{tabular}\\\hline
\end{tabular}
\end{center}
\end{table}

All these form factors share the common property of blowing up in the UV in momentum space, as we shall discuss in Section \ref{sec4}. In general, this implies asymptotic freedom, i.e., interactions are subdominant at short scales.

%%%%%%%%%%%%%%%%%%%%%%%%%%%%%%%%%%%%%%%%%%

\section{The wild beast of nonlocality}

Consider the scalar field theory on Minkowski spacetime
\be
S_\phi = \int \rmd^D x\,\left[\frac12 \phi\B \g(\B)\phi-V(\phi)\right]\,,
\ee
where $V$ is an interaction potential. How could one explore its classical dynamics?

As a first attempt, one can try to truncate the nonlocal operator up to some finite order,
\be
\g(\B)\simeq\sum_{n=0}^{N}c_n\B^n=c_0+c_1\B+\dots+c_N\B^N\,.
\ee
However, the resulting finite-order dynamics is physically inequivalent to the original one and there is no smooth transition between them. The free ($V=0$) case illustrates the point well:
\ba
\g(\B)=\rme^{-l^2\B}&\Rightarrow& \textrm{dispersion relation:}\quad -k^2\rme^{l^2k^2}\phi_k=0\nonumber\\
&\Rightarrow& \textrm{propagator:}\quad-\frac{\rme^{-l^2k^2}}{k^2} \quad\Rightarrow\quad \textrm{1 DOF}\,,\label{dsds}\\
g(\B)\simeq 1-l^2\B&\Rightarrow&\textrm{dispersion relation:}\quad -k^2(1+l^2k^2)\phi_k=0\nonumber\\
&\Rightarrow& \textrm{propagator:}\quad-\frac{1}{k^2}+\frac{1}{l^{-2}+k^2} \quad\Rightarrow\quad \textrm{2 DOF, 1 ghost}\,.
\ea
This example provides a good occasion to comment also on how to determine the number of field degrees of freedom (DOF). In the free case, there is only one double pole, corresponding to 1 DOF. This can also be seen by making a nonlocal field redefinition $\tilde\phi:=\sqrt{\g(\B)}\phi$, so that the free Lagrangian reads $\tilde\cL_\phi=(1/2)\tilde\phi\B \tilde\phi$. This operation is safe if $\g$ is an entire function; if $\g$ is not entire (for instance, $\g=\B^{-n}$), then the field redefinition may result in the elimination of physical modes or the introduction of spurious ones. The Lagrangian $\tilde\cL_\phi$ is second-order in spacetime derivatives and features one local field $\tilde\phi$, hence there is only 1 DOF and solutions are specified by two initial conditions. However, when $V$ is nonlinear of cubic or higher order the field redefinition does not absorb nonlocality completely and one is left with a possibly intractable problem, with extra nonperturbative degrees of freedom \cite{cuta3} and an infinite tower of Ostrogradski modes.

In fact, the Cauchy problem can be na\"ively stated as the assignment of an infinite number of values at some initial time $t_{\rm i}$,
\be
\phi(t_{\rm i}),\,\dot\phi(t_{\rm i}),\,\ddot\phi(t_{\rm i}),\,\dddot\phi(t_{\rm i}),\,\dots\,.
\ee
Thus, paradoxically, we can solve the dynamics only if we already know the solution \cite{MoZ}:
\be\label{tay}
\phi(t)=\sum_{n=0}^{+\infty}\frac{\phi^{(n)}(t_{\rm i})}{n!}(t-t_{\rm i})^n\,.
\ee
If we do not specify all the initial conditions, the solution may be non-unique.

Finally, a word on how to rewrite nonlocalities as a convolution. It is well known that infinitely many derivatives can be traded for integrated kernel functions:
\ba
\phi(x) \g(\B)\,\phi(x) \!\!\!&=&\!\!\! \phi(x)\int\rmd^D k\,\g(-k^2)\,\de(k^\mu-\rmi\N^\mu)\,\phi(x)\nonumber\\
											 &=&\!\!\! \phi(x)\int\rmd^D k\,\left[\int\frac{\rmd^D z}{(2\pi)^D}\,F(z)\,\rme^{-\rmi z^\mu k_\mu}\right]\,\de(k^\mu-\rmi\N^\mu)\,\phi(x)\nonumber\\
											 &=&\!\!\!\phi(x)\int\frac{\rmd^D y}{(2\pi)^D}\,F(y-x)\,\phi(y)\,.
\ea
However, by itself this operator bears no practical advantage. Hiding infinitely many derivatives into integrals does not help in solving the Cauchy problem, \emph{unless} the kernel $F$ could be found by solving some auxiliary, finite-order differential equations. This is precisely the leverage point we will focus on.

%%%%%%%%%%%%%%%%%%%%%%%%%%%%%%%%%%%%%%%%%%

\section{Diffusion method}

The diffusion method was proposed some years ago \cite{cuta2,cuta3,MuNu3,cuta7} to solve nonlocal scalar field theories with exponential form factor \Eq{dsds}, a very specific nonlocal operator that arises in string theory. By trading nonlocal operators with shifts in a fictitious extra direction $r$, the method allows one to count the number of DOF and of initial conditions (which are finite) and to find nonperturbative solutions. All these features can be easily illustrated by the scalar field theory
\be\label{nla}
S_\phi = \int \rmd^D x\,\left[\frac12\phi(x)\B \rme^{-l^2\B}\phi(x)-V(\phi)\right]\,,
\ee
where $l^2$ is a constant. The equation of motion is
\be\label{eoms0}
\B\rme^{-l^2\B}\phi-V'(\phi)=0\,.
\ee

Define now a \emph{localized system}, a priori independent of \Eq{nla}, living in $D+1$ dimensions and featuring two scalars $\Phi(r,x)$ and $\chi(r,x)$:
\ba
\cS[\Phi,\chi]&=&\int \rmd^D x\,\rmd r \left(\cL_{\Phi}+\cL_{\chi}\right)\,,\label{act}\\
\cL_{\Phi}&=&\frac12\Phi(r,x)\B \Phi(r-l^2,x)-V[\Phi(r,x)]\,,\label{locPh2}\\
\cL_{\chi}&=&\frac12 \int_0^{l^2} \rmd q\,\chi(r-q,x)(\p_{r'}-\B)\Phi(r',x),\qquad r'=r+q-l^2.\label{locch2}
\ea
The equations of motion are
\ba
0&=&(\p_r-\B)\Phi(r,x),\qquad 0=(\p_r-\B)\chi(r,x),\\
0&=&\frac12[\B\Phi(r-l^2,x)+\chi(r-l^2,x)]+\frac12[\B\Phi(r+l^2)-\chi(r+l^2)]-V'[\Phi(r,x)]\,.\label{eoms}
\ea
The first line is telling us that the fields are diffusing along the extra direction. At this point, one assumes that there exists a constant $\b$ such that the equation of motion \Eq{eoms} coincides with the one of the nonlocal system \Eq{nla}, Equation\ \Eq{eoms0}, on the slice $r=\b l^2$ (the \emph{physical slice}). This is achieved provided the following conditions hold:
\be\label{exco}
\Phi(\b l^2,x)=\phi(x),\qquad \chi(\b l^2,x) = \B\Phi(\b l^2,x)\,.
\ee

The conclusion is that the localized system has 4 initial conditions $\Phi(r,t_{\rm i},{\bf x})$, $\dot\Phi(r,t_{\rm i},{\bf x})$, $\chi(r,t_{\rm i},{\bf x})$, $\dot\chi(r,t_{\rm i},{\bf x})$ and 2 field DOF $\Phi$ and $\chi$. On the physical slice, because of \Eq{exco} the number of DOF reduces to 1 and the initial conditions on $\chi$ are not independent, so that the initial conditions in the nonlocal theory are on the field $\phi$ and its first derivative:
\be
\phi(t_{\rm i},{\bf x}),\, \dot\phi(t_{\rm i},{\bf x})\,.
\ee
%We can appreciate an exquisite paradox here. When $V'=0$, then $\chi=0$, a condition which reduces both the number of localized DOF and of initial conditions. \emph{The presence or absence of nonlinear interactions $V(\phi)$ determines the number of initial conditions of the nonlocal system.} This result is highly counterintuitive but in line with the concept, typical of nonlocal theories, that the Cauchy problem is determined by the dynamics itself, as we saw when we wrote the Taylor expansion \Eq{tay} around the initial time. While the problem reaches the absurd with Equation\ \Eq{tay}, it becomes understandable and tractable when recast in terms of the higher-dimensional diffusing system.

With traditional methods, only \emph{perturbative} solutions of the linearized EOM or what we call ``static'' (in the extra direction $r$) solutions are available to inspection. By this name, we mean solutions where nonlocality is, in one way or another, trivialized. The most typical approach is to look for field configurations where the field is an eigenfunction of the Laplace--Beltrami operator $\B$:\footnote{In the case of gravity, one has an \emph{Ansatz} of the form $\B R=\la R$ or similar \cite{BMS}.}
\be\label{stati}
\B\phi =\la\phi\quad\Rightarrow\quad \rme^{-l^2\B}\phi=\rme^{-l^2\la}\phi\,.
\ee
In contrast, the diffusion method gets access to nonperturbative solutions valid in the presence of nonlinear interactions and nontrivial nonlocality. These solutions are, in general, only approximate, and are found by searching for the value of $\b$ minimizing the equations of motion. Here is a list of solutions obtained with this method:
\begin{itemize}
\item $\phi(t)={\rm Kummer}(t)$ on a Friedmann--Robertson--Walker (FRW) background \cite{cuta2}. 
\item Rolling tachyon $\phi(t)=\sum_n a_n\rme^{nt}$ in string field theory, $V=\phi^3$ \cite{roll}.
\item Kink $\phi(x)={\rm erf}(x)$, $V=\phi^3$ \cite{cuta4,cuta5,cuta7}.
\item $\phi(t)=\gamma(\a,t)$ (incomplete gamma function), $V=\phi^n$ on FRW background \cite{cuta5,cuta6}.
\item Instanton $\phi(x)={\rm erf}(x)$, $V=\phi^4$ (brane tension recovered at 99.8\% level) \cite{cuta5}.
\item Kink $\phi(x)={\rm erf}(x)$, $V=(\rme^\B\phi^2)^2$ \cite{cuta5}.
\item Various profiles $\phi(t)$ in bouncing and singular cosmologies  \cite{cuta8}.
\end{itemize}
The main reason why diffusion works is that nonlocal operators are represented as a shift in an extra direction rather than as an infinite sum of derivatives. The latter representation does not span the whole space of solutions, as one can see by a toy example \cite{cuta2}. Consider a $D=4$ FRW background with Hubble expansion $H=\dot a/a=H_0/t$, the Laplace--Beltrami operator $\B=-\p_t^2-3H\p_t$, and the homogeneous power-law profile $\phi(t)=t^p$. If we try to calculate the object $\rme^{r\B}\phi$ as a series, the result diverges: $\rme^{r\B}\phi=\sum_{n=0}^\infty (r\B)^n\phi/n! =\infty$. On the other hand, with the diffusion method one interprets $\phi(t)=\Phi(t,0)$ as the initial condition in the diffusion scale $r$ and the profile $\rme^{r\B}\Phi(t,0)={\rm Kummer}(t,r)$ is a linear superposition of well-defined Kummer functions.

%%%%%%%%%%%%%%%%%%%%%%%%%%%%%%%%%%%%%%%%%%

\section{Initial conditions and degrees of freedom}

The diffusion method has been extended to the case of gravity in \cite{CMN2} for the string-related and Krasnikov exponential form factors and in \cite{CMN3} for the asymptotically polynomial (Kuz'min and Tomboulis--Modesto) form factors. The reader can consult those papers for technical details; here we only quote the bottom line, which is that, for the string-related form factor, the localized system associated with \Eq{action} has 6 initial conditions $g_{\mu\nu}(t_{\rm i},{\bf x})$, $\dot g_{\mu\nu}(t_{\rm i},{\bf x})$, $\Phi_{\mu\nu}(r,t_{\rm i},{\bf x})$, $\dot\Phi_{\mu\nu}(r,t_{\rm i},{\bf x})$, $\chi_{\mu\nu}(r,t_{\rm i},{\bf x})$, $\dot\chi_{\mu\nu}(r,t_{\rm i},{\bf x})$, two for each rank-2 symmetric tensor field (the metric $g_{\mu\nu}$ and the tensors $\Phi_{\mu\nu}$ and $\chi_{\mu\nu}$). Since, on the physical slice, $\Phi_{\mu\nu}(\b l^2,x)=G_{\mu\nu}$ and $\chi_{\mu\nu}(\b l^2,x)=R_{\mu\nu}$, solutions of the nonlocal system \Eq{action} are characterized by 4 initial conditions:
\be
g_{\mu\nu}(t_{\rm i},{\bf x}),\,\dot g_{\mu\nu}(t_{\rm i},{\bf x}),\,\ddot g_{\mu\nu}(t_{\rm i},{\bf x}),\,\dddot g_{\mu\nu}(t_{\rm i},{\bf x})\,.
\ee

Regarding the degrees of freedom, the counting for the exponential form factor is the following.
\begin{itemize}
\item Graviton $g_{\mu\nu}$: symmetric $D\times D$ matrix with $D(D+1)/2$ independent entries, to which one subtracts $D$ Bianchi identities $\N^\mu G_{\mu\nu}=0$ and $D$ diffeomorphisms (the theory is fully diffeomorphism invariant). Total: $D(D-3)/2$. In $D=4$, there are 2 degrees of freedom, the usual polarization modes.
\item Tensor $\phi_{\mu\nu}$: symmetric $D\times D$ matrix with $D(D+1)/2$ independent entries, to which one subtracts $D$ transverse conditions $\N^\mu \phi_{\mu\nu}=0$. Total: $D(D-1)/2$. In $D=4$, there are 6 degrees of freedom.
\item Grand total: $D(D-2)$.
\end{itemize}
In $D=4$, there are 8 DOF. Two of them (the graviton) are visible already at the perturbative level, while the other 6 are of nonperturbative origin. Their role in phenomenology \cite{CMN3} remains to be determined. 

%%%%%%%%%%%%%%%%%%%%%%%%%%%%%%%%%%%%%%%%%%

\section{Quasi-uniqueness in nonlocal gravity}\label{sec4}

Similar results hold for the asymptotically polynomial form factors, although in that case the diffusion method requires some adjustment. We present here a longer derivation than the one given in \cite{CMN3}. Any form factor can be written in terms of a kernel $\cG$ governed by a simple system of renormalization-group-like differential equations, which determine how $\cG$ varies in the \emph{space of all possible functionals $P(\B)$}. To parametrize this space, we replace $P(\B)$ with $r P(\B)$ in Equation\ \Eq{Hz}, where $r$ is a dimensionless parameter eventually sent to 1 (end of the flow):
\be
H_r(\B) := \a\int_0^{r P(\B)} \rmd\omega\, \frac{1-f(\omega)}{\omega}\,\,\stackrel{r\to 1}{\longrightarrow}\,\, H(\B)\,.\label{Hz2}
\ee
In particular, we can write the form factor $\exp H$ as
\be
\rme^{H_r(\Box_x)} {\mathcal{R}}(x) = \int \rmd^D y \sqrt{-g} \, \cG(x,y;r)\,{\mathcal{R}}(y)\,,\label{ExplicitNL}
\ee
where $\cG(x,y;r)$ is a kernel formally given by
\be
\cG(x,y;r) = \rme^{H_r(\B_x)} \cG(x,y;0)= \rme^{H_r(\B_x)}\frac{\delta^D(x-y)}{\sqrt{-g(y)}},\label{Gsol}
\ee
and ${\mathcal R}$ is a generalized curvature: the Ricci scalar, the Ricci tensor, or the Riemann tensor. Notice that $\nabla_\mu g_{\nu\rho} = 0$, hence the form factor does not act upon the determinant of the metric. Equation \Eq{Gsol} has two features. First, it carries infinitely many derivatives. Second, its nonlocality is especially difficult to deal with because in quantum gravity the form factor $\exp H$ explodes in Euclidean momentum space, so that its inverse (the propagator) is suppressed in the UV. For instance, for Kuz'min Euclidean form factor, $H_r(-k^2)=\a[\ln(l^2k^2)+\Gamma(0,l^2k^2)+\gamma_{\rm E}]\to+\infty$ as $|k|\to\infty$; for the string form factor, $H_r(-k^2)=\a r l^2 k^2$; and so on. Therefore, Equation\ \Eq{Gsol} cannot be calculated directly for $r\geq 0$ in all realistic cases. To address these issues, we can follow two roads, both of which give up Equation\ \Eq{Gsol}. One is to look for regularized solutions \cite{roll} using the diffusion method coupled with a wave equation \cite{cuta7}, but this strategy may be too complicated for the asymptotically polynomial case. Another way out is to consider the inverse problem
\ba
\cF(x,y;r) &=& \rme^{-H_r(\B_x)} \cF(x,y;0)\,,\label{Fsol1}\\
\cF(x,y;0) &=& [-g(y)]^{-1/2}\de^D(x-y)\,,\label{Fsol2}
\ea%\es
where $\cF=\cG^{-1}$ and now the nonlocal operator is damped at high energies. Once $\cF$ is found, one can determine $\cG=\rme^{2H_r}\cF$ %(possibly through regularization \cite{roll})
 with deconvolution methods \cite{Ulm10}. This reformulation does not solve the infinite-derivative issue, which is addressed by replacing \Eq{Fsol1} and \Eq{Fsol2} with the following finite-order differential equations for two kernels $\cK$ and $\cF$ with specified boundary conditions in the diffusion parameter $r$:
\bs\label{MasterSystem}\ba
&&r \partial_r \cF(x,y;r) = \a [(\cK \star \cF)(x,y;r)-\cF(x,y;r)]\,,\label{Master2}\\
&&(\cK \star \cF)(x,y;r):= \int \rmd^D x' \sqrt{-g}\, \cK(x,x';r) \, \cF(x',y;r)\,,
\ea
where $\cK$ is the kernel associated with the operator $f[r P(\B)]$. The solution of Equation\ \Eq{Master2} is formally given by \Eq{Fsol1}. For the general class $f(\om)=\exp(-\om)$ (asymptotically polynomial $H^{\rm pol}_r$, Equation\ \Eq{FFgen} with $P\to r P$; includes Kuz'min and Tomboulis--Modesto form factors), we also have
\be
\left[\p_r+P(\Box_x)\right] \cK(x,y;r) = 0,\qquad\cK(x,y;0)=\cF(x,y;0).\label{Master1}
\ee\es
Note that Equation\ \Eq{Master1} corresponds, in quantum gravities, to diffusion in the correct direction and its solution $\cK(x,y;r) = \rme^{-r P(\Box_x)} \cK(x,y;0)$ is well defined.

In the much simpler case of the general class $f(\om)=1-\om$ (exactly monomial $H^{\rm exp}_r$, Equation\ \Eq{FFgen2} with $P\to r P$; includes the string-related and Krasnikov exponential form factors), Equations\ \Eq{MasterSystem} are replaced by just one equation:
\be\label{MasterSystem2}
[\p_r+\a P(\Box_x)]\cF(x,y;r)=0\,,
\ee
whose solution $\cF(x,y;r) =\rme^{-\a r P(\B_x)}\cF(x,y;0)$ is well defined and coincides with \Eq{Fsol1}. 

The form factor $\exp H$ is not the only nonlocal operator in the action (\ref{action}). We also need to specify $\Box^{-1}$ by means of the Green function solving \cite{DeWo2}
\be
\Box \tilde{G}(x,y) = [-g(y)]^{-1/2}\de^D(x-y)\,.\label{DW1}%\frac{\delta^D(x-y)}{\sqrt{-g(y)}}
\ee

The equations of motion can be expressed in terms of the kernels $\cK(x,x';r)$ and $\cF(x,y;r)$ and the Green function $\tilde{G}(x,y)$. These kernels are nonlocal because they depend on two spacetime points, but they are determined by the above sets of master equations independent of the actual equations of motion. The system is of finite differential order, which leads to a finite number of initial conditions. In particular, when $P(\B)=P_1(\B)$ we have second-order differential equations for the kernels. This might be regarded as a criterion to select the string-related and Kuz'min form factors among all the possibilities. In this way, the plethora of nonlocal quantum gravities is reduced to only two choices.

%%%%%%%%%%%%%%%%%%%%%%%%%%%%%%%%%%%%%%%%%%

\section{Discussion}\label{disc}

In this paper, we reviewed and expanded upon recent results on the number of initial conditions and degrees of freedom in some nonlocal theories of gravity \cite{CMN2,CMN3}. A question left open is whether the diffusion method covers all solutions of the nonlocal system.  All solutions of nonlocal systems found to date (which are not many) are of four types:
\begin{enumerate}
\item[(i)] Exact and in common with Einstein gravity \cite{LMRa,CMM} such as the Schwarzschild metric and any other Ricci-flat solution ($R_{\mu\nu}=0$).
\item[(ii)] Exact of ``static'' type \Eq{exco}, i.e., $r$-independent and describing trivial diffusion \cite{BMS,KSKS}. Static solutions are eigenfunctions of the Laplace--Beltrami operator $\B$.
\item[(iii)] Diffusion solutions as those described in this paper. Diffusing solutions are usually approximate with a high degree of accuracy, although exact solutions of scalar-field toy models also exist.
\item[(iv)] Approximate solutions of linearized equations, which are much better known than the other types, both in astrophysics \cite{Tse95,FZdP,Fro15,FrZe1,Fro16,FrZe3} and cosmology \cite{CMNi}.
\end{enumerate}
Types (i)--(iii) hold for \emph{nonlinear} EOM with \emph{interactions}, while (iv) are perturbative and valid only for linearized EOM. The relation between solutions of the nonlinear and linear dynamics is not completely clear, since there is no uniqueness theorem (such as Birkhoff's) about backgrounds with symmetries. Therefore, we can only compare (i)--(iii) among one another. Obviously, diffusing solutions (iii) encompass (i) and (ii), which are both trivial in $r$; the diffusion method is not necessary there. It is also highly likely (but we have no conclusive proof of it) that \emph{all} solutions are of diffusing type. The argument goes as follows. On one hand, the background-independent exact EOM are recovered with the diffusion method. On the other hand, this happens only because we have represented the nonlocal operator $\exp\B$ as a shift in the fake direction $r$, instead of as a series in $\B^n$. However, the representation in terms of series does not cover the whole space of solutions spanned by the representation in terms of a diffusing direction \cite{cuta2}. Also, as we saw, the representation of nonlocal operators as integral kernels is quite general. Therefore, we expect the diffusion method\footnote{The kernels that appear in nonlocal quantum gravity obey a set of master equations among which there is a diffusion equation. This is the reason why we still call the generalized procedure of Section \ref{sec4} diffusion method.} to span all solutions. At least, we cannot find any convincing argument why, or any counter-example in the literature showing that, this should not be the case. A counter-example would be a solution $\phi(x)=\Phi(\b l^2,x)$ such that the function $\Phi(r,x)$ does not obey a diffusion equation in $r$.

Regardless of the validity of this conjecture, we believe that the diffusion method will be a very useful tool to explore the nonlinear dynamics of nonlocal gravity. The fact that we have been able to limit the choice of form factors simply by imposing that the dynamical and master equations are of finite order is already an advance, but other physical applications are possible. Clearly, being able to find solutions to the classical equations will be important to study the phenomenology produced by these theories, in particle-physics experiments as well as in cosmological and astrophysical observations. In particular, the fate of cosmological and astrophysical singularities (big bang and black holes) can be determined by solving the fully nonlinear equations of motion, thus going beyond the approximations and assumptions usually employed.

%%%%%%%%%%%%%%%%%%%%%%%%%%%%%%%%%%%%%%%%%%
\vspace{6pt} 

%%%%%%%%%%%%%%%%%%%%%%%%%%%%%%%%%%%%%%%%%%

\funding{The author is supported by the I+D grants FIS2014-54800-C2-2-P and FIS2017-86497-C2-2-P.}

%%%%%%%%%%%%%%%%%%%%%%%%%%%%%%%%%%%%%%%%%%
\conflictsofinterest{The author declares no conflict of interest.} 

\reftitle{References}

\end{document}